\documentclass[prl,aps,twocolumn,tightenlines,superscriptaddress]{revtex4}
\usepackage{amsmath}
\usepackage{graphicx}
\usepackage{color}
\usepackage{amssymb}

\begin{document}

\title{Large Tensor-to-Scalar Ratio in Small-Field Inflation}

\author{Takeshi Kobayashi}
\affiliation{%
Canadian Institute for Theoretical Astrophysics,
University of Toronto, 60 St. George Street, Toronto, Ontario M5S
3H8, Canada}
\affiliation{%
Perimeter Institute for Theoretical Physics, 
31 Caroline Street North, Waterloo, Ontario N2L 2Y5, Canada
}

\author{Tomo Takahashi}
\affiliation{Department of Physics, Saga University, Saga 840-8502, Japan}


\begin{abstract}
We show that density perturbations seeded by the inflaton 
 can be suppressed when having additional light degrees of freedom contributing to
 the production of perturbations. The inflaton fluctuations
 affect the light field dynamics by modulating the length of the
 inflationary period, hence produce additional density perturbations in
 the post-inflationary era. Such perturbations can cancel those
 generated during inflation as both originate from the same inflaton
 fluctuations. This allows production of large
 gravitational waves from small-field inflation, which is normally
 forbidden by the Lyth bound on the inflaton field excursion. We also
 find that the field bound is taken over by the light scalar when the
 inflaton-induced perturbations are suppressed, thus present a
 generalized form of the Lyth bound that applies to the total field
 space. 
\end{abstract}


\maketitle

{\it Introduction.}--- The origin of cosmic structure, or primordial
density perturbations, is an important issue in cosmology. One of
the key observables is the gravitational waves which are usually quantified by
the tensor-to-scalar ratio $r$.
In standard inflationary scenarios,
there exists the so-called Lyth bound~\cite{Lyth:1996im}
which constrains~$r$ in terms of the inflaton field
excursion as $r \lesssim 0.01 (\Delta \phi / M_p)^2$ 
with $M_p$ being the reduced Planck mass.
This can be rephrased as small-field (i.e. sub-Planckian) inflation
models only producing tiny~$r$. 
The bound is due to the inflaton with super-horizon field fluctuations
necessarily producing density perturbations, and it should be noted
that the bound can become more restrictive, but not
alleviated, by non-slow-roll inflation (without superluminal
modes)~\cite{Baumann:2011ws}, or by simply adding extra sources
for the density perturbations~\cite{Linde:2005he}. 

In this letter we point out that this general belief is not
necessarily true when there are additional light fields producing density
perturbations such as in the curvaton scenario~\cite{Enqvist:2001zp} or
modulated reheating~\cite{Dvali:2003em}. 
The basic idea can be explained as follows: The inflaton field
fluctuations source density perturbations by giving slightly longer/shorter
inflationary periods among different patches of the universe.
This also affects the dynamics of the light fields by allowing more/less
time to roll along their potentials during inflation. Therefore 
the inflaton field fluctuations induce fluctuations of the light fields
as well, leading to further generation of density perturbations in the
post-inflationary era. The inflaton-induced perturbations generated
during and after inflation (note that both are seeded by the same
inflaton fluctuations) 
can cancel each other, then one can evade the Lyth bound for the
inflaton. We demonstrate that density perturbations from the
inflaton fluctuations can actually be suppressed, and show that large
tensor-to-scalar ratio can be obtained even in small-field inflation
models with the aid of an additional light field generating perturbations.  

However, since the total perturbations generated by the two fields
remain finite, we will also see that in order to evade
the field range bound for the inflaton, the other field instead has to
take over the large field excursion. In this sense the Lyth bound is
shown to apply to the total field space of the inflaton and the light field.
We will derive a generalized form of the Lyth bound, then present an
explicit scenario where perturbations from the inflaton are inevitably
suppressed.

\medskip
{\it Lyth bound for the inflaton and a light scalar.}---
Let us study field range bounds when there is an extra light scalar~$\sigma$
besides the inflaton~$\phi$ producing density perturbations as in, for
e.g., the curvaton or modulated reheating mechanisms.  
We assume that the $\sigma$ field has negligibly tiny energy density
during inflation and also that its dynamics has little effect on the
inflationary expansion, but contributes to the production of density
perturbations in the post-inflationary era. 
The fields $\phi$ and $\sigma$ interact
with each other only via gravity.

In order to compute the density perturbations seeded by the fields'
fluctuations, we use the $\delta \mathcal{N}$-formalism
and study the evolution of the universe which is uniquely determined by a
set of field values $(\phi, \, \sigma)$ at an arbitrary time when the
fields follow attractor solutions. 
The density perturbations are obtained by computing
the fluctuations in the e-folding number:
\begin{equation}
 \mathcal{N} = 
\int_{t_*}^{t_{\mathrm{end}}} H dt +
\int_{t_{\mathrm{end}}}^{t_{\mathrm{f}}} H dt 
 \equiv \mathcal{N}_a + \mathcal{N}_b,
\end{equation}
where the subscript ``$*$'' denotes quantities when the CMB scale
exits the horizon, ``end'' at the end of inflation, and ``f'' at the final
uniform density hypersurface after which no further $\delta \mathcal{N}$
is produced. $\mathcal{N}$ is split into $\mathcal{N}_a$ and $\mathcal{N}_b$ at
$t_{\mathrm{end}}$. Considering the fields to follow attractor solutions at
least until the end of inflation, then
\begin{gather}
 \frac{\partial \mathcal{N}}{\partial \phi_*} = 
 \frac{\partial \mathcal{N}_a}{\partial \phi_*} +
 \frac{\partial \phi_{\mathrm{end}}}{\partial \phi_*}
 \frac{\partial \mathcal{N}_b}{\partial \phi_{\mathrm{end}}}  + 
 \frac{\partial \sigma_{\mathrm{end}}}{\partial \phi_*} 
 \frac{\partial \mathcal{N}_b}{\partial \sigma_{\mathrm{end}}},
\\
  \frac{\partial \mathcal{N}}{\partial \sigma_*} = \frac{\partial
  \sigma_{\mathrm{end}}}{\partial \sigma_*} 
 \frac{ \partial \mathcal{N}_b}{\partial \sigma_{\mathrm{end}}}.
\end{gather}
Note that $\mathcal{N}_a $ and $\phi_{\mathrm{end}}$ are independent
of~$\sigma_*$ since the inflationary expansion is governed
by the inflaton.  
Hereafter we suppose that the end of inflation is set by a constant
inflaton field value~$\phi_{\mathrm{end}}$.
(Hence we do not consider inhomogeneous end of
inflation~\cite{Bernardeau:2002jf}, but the generalization to such cases is
straightforward.) This yields
\begin{equation}
 \frac{\partial \mathcal{N}_a}{\partial \phi_*} 
= \frac{\partial }{\partial \phi_*} 
 \int_{\phi_*}^{\phi_{\mathrm{end}}} \frac{H}{\dot{\phi}} d\phi 
=  - \frac{H_*}{\dot{\phi}_*},
\end{equation}
where an overdot represents a time derivative.

Considering the field~$\sigma$ to be light during inflation and that it
slow-rolls along its potential~$V(\sigma)$, i.e. $ 3 H \dot{\sigma} = -
V'(\sigma) $ with a prime denoting a $\sigma$-derivative, one obtains
\begin{equation}
 \int^{\sigma_{\mathrm{end}}}_{\sigma_*} \frac{d \sigma }{V'(\sigma)} = 
 - \int^{\phi_{\mathrm{end}}}_{\phi_*}\frac{d\phi }{3 H \dot{\phi}}.
\label{eq6}
\end{equation}
Partially differentiating both sides by $\phi_*$ gives
\begin{equation}
 \frac{\partial \sigma_{\mathrm{end}}}{\partial \phi_*} =
  \frac{V'(\sigma_{\mathrm{end}})}{3 H_* \dot{\phi}_*},
\end{equation}
while differentiating with $\sigma_*$ (note that 
the right hand side of (\ref{eq6}) is independent of~$\sigma_*$)
yields
\begin{equation}
 \frac{\partial \sigma_{\mathrm{end}}}{\partial \sigma_*} = 
 \frac{V'(\sigma_{\mathrm{end}})}{V'(\sigma_*)}.
\end{equation}
Hence by combining the above results, one arrives at
\begin{equation}
 \frac{\partial \mathcal{N}}{\partial \phi_*}  = -\frac{H_*}{\dot{\phi}_*} 
 ( 1 - \kappa  ),
\quad \mathrm{where} \quad
 \kappa \equiv 
\frac{V'(\sigma_*)}{3 H_*^2} \frac{\partial \mathcal{N}}{\partial
 \sigma_*} .
\label{kappa}
\end{equation}
This $\kappa$ represents the effect of $\delta \phi$ modulating the
field value of~$\sigma$ at the end of inflation, and thus further
generating density perturbations in the post-inflationary era. 

Let us suppose Gaussian field fluctuations with power spectra
$\mathcal{P}^{1/2}_{\delta \phi_*} = \mathcal{P}^{1/2}_{\delta \sigma_*}
= H_* / 2 \pi$, with no correlations between the two. Then the total
density perturbations are written as a sum of contributions from each field,
\begin{equation}
 \mathcal{P}_\zeta = \mathcal{P}_{\zeta \phi} + \mathcal{P}_{\zeta
  \sigma}
 = 
\left( \frac{\partial \mathcal{N}}{\partial \phi_*} \frac{H_*  }{2 \pi} \right)^2 + 
\left( \frac{\partial \mathcal{N}}{\partial \sigma_*} \frac{H_*  }{2 \pi} \right)^2.
\label{eq12}
\end{equation}
The amplitude of tensor perturbations
(gravitational waves)~$\mathcal{P}_T$ is set only by the inflation scale, 
so in terms of the tensor-to-scalar ratio~$r$ defined as 
\begin{equation}
 r \equiv \frac{\mathcal{P}_T}{\mathcal{P}_\zeta }
\quad \mathrm{with} \quad
 \mathcal{P}_T = \frac{2 H_*^2}{\pi^2 M_p^2} ,
\end{equation}
we arrive at the generalized form of the Lyth bound:
\begin{align}
 \left( \frac{1}{M_p}\frac{\dot{\phi}_*}{H_*}   \right)^2
 & = \frac{(1 - \kappa )^2}{8}\frac{\mathcal{P}_\zeta
 }{\mathcal{P}_{\zeta \phi} }  r 
 \geq \frac{(1 - \kappa)^2 }{8} r,
\label{Lythphi}
 \\
 \left( \frac{1}{M_p}\frac{\dot{\sigma}_*}{H_*}   \right)^2
 & = \frac{\kappa^2}{8}\frac{\mathcal{P}_\zeta }{\mathcal{P}_{\zeta \sigma} }  r 
 \geq \frac{\kappa^2  }{8}  r.
\label{Lythsigma}
\end{align}
The left hand sides represent the field excursions during one Hubble time
at around when the CMB scale exits the horizon. The total field
excursions can be estimated by multiplying the expressions by the total number
of inflationary e-folds \footnote{The total field excursion would not simply be  
(\ref{Lythphi}) multiplied by the total e-folding number if $d\phi /
d\mathcal{N}$ suddenly changes during inflation~\cite{BenDayan:2009kv}. 
However, even in such cases, obtaining observably large~$r$ over the
entire CMB scales from a single inflaton field requires $\Delta \phi$ of
order the Planck scale, if not super-Planckian.}.
One recovers the familiar result for a single inflaton for $\kappa = 0$,
i.e. when there is no additional degree of freedom generating density perturbations.

On the other hand when $\kappa \approx 1$,
density perturbations sourced from the inflaton
fluctuations are suppressed (cf. (\ref{kappa})), 
and the lower bound on the inflaton field excursion~(\ref{Lythphi})
becomes tiny. However it should be noted that in such cases, the required
field range for~$\sigma$ (\ref{Lythsigma}) increases instead.
Thus one sees that the familiar form of the Lyth bound now applies to
the total field space of $\phi$ and $\sigma$. 

One can further check that the total power spectrum (9) can be nearly
scale-invariant when both $\phi$ and $\sigma$ slow roll during
inflation.

Before ending this section, we should also remark on the variation of
the energy density of the field~$\sigma$:
\begin{equation}
\left|\frac{1}{3 M_p^2 H_*^2} \frac{\dot{V}(\sigma_*)}{H_*} \right|
 =  \frac{\kappa^2}{8} \frac{\mathcal{P}_{\zeta}}{\mathcal{P}_{\zeta
 \sigma}} r
 >  \frac{\kappa^2}{8}r ,
\end{equation}
which is similar to the field bound (\ref{Lythsigma}) except for the
power of the left hand side.
One sees that $\kappa^2 r $ as large as $\sim 0.1$ requires 
the variation $\Delta V(\sigma)$ during inflation to be comparable to
the inflation scale itself. In such a case $\sigma$ is also regarded as
the inflaton in the sense that its dynamics directly affects the
inflationary expansion, then the discussions in this section need to be
modified accordingly.

\begin{figure}[bp]
 \includegraphics[width=.8\linewidth]{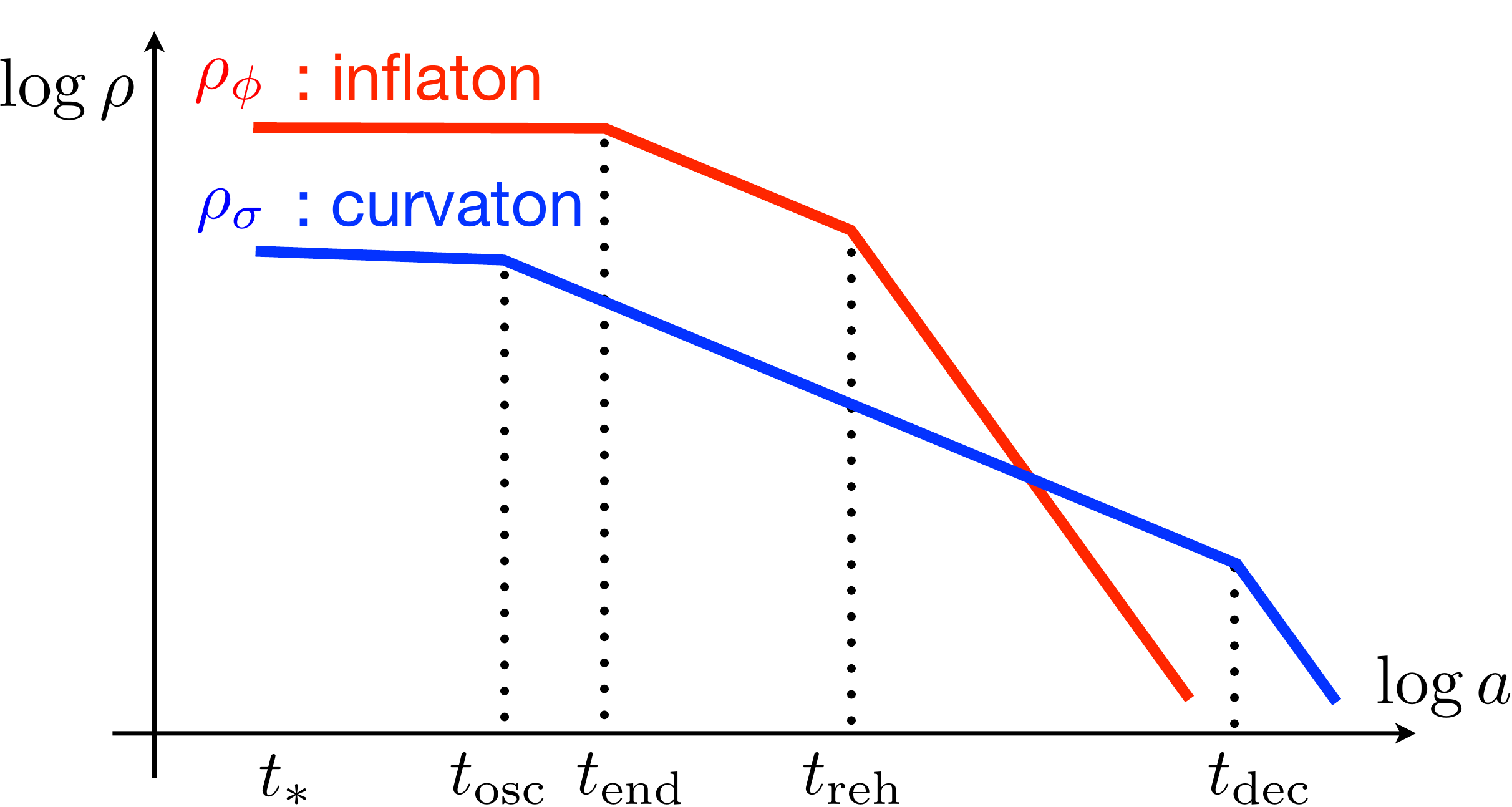}
 \caption{Schematic of the time variation of energy densities.}
 \label{fig:schematic}
\end{figure}

\medskip
{\it An example: early oscillating curvaton.}---
As an example, we present a curvaton scenario where the
effective potential forces the curvaton to start oscillating during
inflation (cf. Fig.~\ref{fig:schematic}).
In such cases, inflaton-induced perturbations 
are inevitably suppressed given that the curvaton dominates the
universe before decaying away. 
The computations in the previous section do not strictly apply here
since the curvaton ceases to slow-roll during inflation
(cf. discussions around (\ref{eq6})), however the basic assumptions and
arguments stay the same \footnote{The inflaton-induced perturbations can also 
be suppressed in the conventional curvaton scenario where the oscillation
starts after inflation. Though we remark that for a simple
quadratic curvaton potential $V =  m_\sigma^2 \sigma^2$/2, the
parameter~$\kappa$ is suppressed by $m_\sigma^2 / H_*^2$ and is much
smaller than unity. One also finds that in order to have $\kappa
\approx 1$ from modulated reheating, then the time variation of the
inflaton decay rate becomes rather large, and thus 
treatments beyond sudden-decay approximations may be
required~\cite{Kobayashi:toappear}.}.  

We consider a curvaton~$\sigma$ whose potential~$V(\sigma)$ allows the
field to slow-roll while the CMB scales exit the horizon, but starts the
curvaton oscillation before the end of inflation when $\sigma$
approaches~$\sigma_{\mathrm{osc}}$. 
Here, $\sigma_{\mathrm{osc}}$ is a constant field value that is
basically set by~$V(\sigma)$ \footnote{One may wonder whether
$\sigma_{\mathrm{osc}}$ is homogeneous since it may
also depend on the Hubble parameter, 
as is seen in the condition for the onset of oscillation 
$|\dot{\sigma}/ H  |_{\mathrm{osc}} \sim | \sigma_{\mathrm{osc}} \! - \!
\sigma_{\mathrm{min}}  |$ where $\sigma_{\mathrm{min}} $ is the
potential minimum~\cite{Kawasaki:2011pd}. We note that since $H$ is
nearly constant during inflation, $\sigma_{\mathrm{osc}}$ can be treated 
as a homogeneous constant.}.
An explicit example of such potential is given later in (\ref{linearpot}). 
$V(\sigma)$ is assumed to be well approximated by a quadratic around
its minimum, and thus the energy density of the oscillating curvaton
redshifts as nonrelativistic matter \footnote{The curvaton may
initially undergo nonsinusoidal oscillations (e.g. (\ref{linearpot})),
but this does not alter the main results.}. 

Inflation ends when $\phi$ approaches a constant field
value~$\phi_{\mathrm{end}}$, after which the inflaton also behaves as
matter until it decays into 
radiation (reheating), and eventually the curvaton also decays away and thermalize
with the inflaton decay products. 
We assume the fields to suddenly
decay when $H$ is equal to their constant decay rates $\Gamma_\phi$ and
$\Gamma_\sigma$. The curvaton energy density is 
negligibly tiny during inflation, hence also until reheating. 

Taking the final hypersurface as when the curvaton decays, let us
break up the e-folding number as
(the subscript ``osc'' denotes the onset of curvaton oscillation,
``reh'' the inflaton decay, and ``dec'' the curvaton decay)
\begin{equation}
\begin{split}
\mathcal{N} &= \left(
\int^{t_{\mathrm{osc}}}_{t_*} + 
\int^{t_{\mathrm{end}}}_{t_{\mathrm{osc}}} + 
\int^{t_{\mathrm{reh}}}_{t_{\mathrm{end}}} + 
\int^{t_{\mathrm{dec}}}_{t_{\mathrm{reh}}} 
\right) Hdt \\
& \equiv 
 \mathcal{N}_a + \mathcal{N}_b + \mathcal{N}_c + \mathcal{N}_d.
\end{split}
\end{equation}
Then we have
\begin{equation}
 \frac{\partial }{\partial \phi_*} \left( \mathcal{N}_a + \mathcal{N}_b
   \right) = - \frac{H_*}{\dot{\phi}_*} , 
\quad 
 \frac{\partial }{\partial \sigma_*} \left( \mathcal{N}_a + \mathcal{N}_b
				   \right) = 0.
\label{NaNbphi}
\end{equation}
Since $H_{\mathrm{end}}$ is set by the constant~$\phi_{\mathrm{end}}$,
the e-folding number
$\mathcal{N}_c = \ln (H_{\mathrm{end}}/\Gamma_\phi)^{2/3}$
is independent of $\phi_*$ or $\sigma_*$. 
Denoting the energy density of radiation produced from the inflaton decay
by $\rho_r$, then $\dot{\rho}_r = - 4 H \rho_r$ gives
\begin{equation}
 \mathcal{N}_d = \frac{1}{4} \ln \frac{\rho_{r\,
  \mathrm{reh}}}{\rho_{r\, \mathrm{dec}}} =
\frac{1}{4} \ln \frac{3 M_p^2 \Gamma_\phi^2}{3 M_p^2 \Gamma_\sigma^2 -
\rho_{\sigma\, \mathrm{dec}}},
\label{N_d}
\end{equation}
with the curvaton's energy density upon its decay 
\begin{equation}
 \rho_{\sigma\, \mathrm{dec}}
 =
 V(\sigma_{\mathrm{osc}}) \exp
 \left\{ -3 (\mathcal{N}_b + \mathcal{N}_c + \mathcal{N}_d)  \right\}.
 \label{rho_sigmadec}
\end{equation}
Partially differentiating both sides of (\ref{N_d}) yields
\begin{equation}
 \frac{\partial \mathcal{N}_d}{\partial \phi_*} =
 - \frac{3 \hat{r}}{ 4 + 3 \hat{r} }
\frac{\partial
 \mathcal{N}_b}{\partial \phi_*}  ,
\label{benri}
\end{equation}
where $\hat{r}$ is the energy density ratio at curvaton decay,
\begin{equation}
 \hat{r} \equiv \left. \frac{\rho_\sigma}{\rho_r}
		\right|_{\mathrm{dec}}.
\end{equation}
We note that (\ref{benri}) also holds for $\partial / \partial \sigma_*$. 

Now let us consider the time variation of H during inflation to be small,
and adopt a constant scale~$H_{\mathrm{inf}}$ to represent the
inflationary Hubble parameter, i.e. $H \simeq H_{\mathrm{inf}}$. 
(This treatment will induce errors of $ \sim |\dot{H}/H^2|$.) 
Then the slow-roll approximation for the curvaton gives 
\begin{equation}
 \mathcal{N}_a \simeq \int^{\sigma_*}_{\sigma_\mathrm{osc}} d\sigma \, 
 \frac{3 H_{\mathrm{inf}}^2}{ V'(\sigma)},
\label{Naeo}
\end{equation}
which leads to
\begin{equation}
  \frac{\partial \mathcal{N}_a}{\partial \phi_*} \simeq 
 \mathcal{O} \biggl( \frac{\dot{H}}{H^2}  \biggr) \times
 \frac{H_*}{\dot{\phi}_*}, 
\quad \, 
 \frac{\partial \mathcal{N}_a}{\partial \sigma_*} \simeq
 \frac{3 H_{\mathrm{inf}}^2}{V' (\sigma_*)} .
\label{eq:27}
\end{equation}
The $\phi_*$-dependence drops out of the approximation (\ref{Naeo}),
hence we have given an order of magnitude estimation for
$\partial \mathcal{N}_a / \partial \phi_*$. 
Therefore, by combining the above equations, we arrive at the final results
\begin{gather}
 \frac{\partial \mathcal{N}}{\partial \phi_*}  \simeq
 -\frac{4}{4 + 3 \hat{r} } \frac{H_*}{\dot{\phi}_*} ,
\label{Nphieo}
\\
 \frac{\partial \mathcal{N}}{\partial \sigma_*} \simeq
 \frac{3 \hat{r}}{4 + 3\hat{r}} \frac{3 H_{\mathrm{inf}}^2}{
 V'(\sigma_*)}
 \simeq -  \frac{3 \hat{r}}{4 + 3\hat{r}} 
 \frac{H_{*}}{\dot{\sigma}_*}
\label{Nsigmaeo}
\end{gather}
We repeat that these expressions contain errors of order $|\dot{H}/H^2|$
during inflation (see also discussions at the end of this
section) \footnote{Other approximations such as the slow-roll of
$\sigma$ and the neglect of $\rho_\sigma$ during inflation can also
induce errors. Furthermore, the behavior of the oscillating 
curvaton deviates from that of matter at time scales shorter than
its oscillation period. Such effects give rise to corrections 
typically of $\sim H_{\mathrm{inf}}/m_\sigma$ ($m_\sigma$: mass during
oscillation).}\footnote{Since we have  
used the approximation $H\simeq H_{\mathrm{inf}}$, 
one should be careful when discussing scale-dependencies of the
perturbations. In particular, one should not simply carry out a
$k$-derivative of the far right hand side of (\ref{Nsigmaeo}) when
computing the spectral index.}.
Here one clearly sees that perturbations from the inflaton are highly
suppressed if the curvaton dominates the universe before decaying,
i.e. $\hat{r} \gg 1$. 
The suppression of the inflaton-induced perturbations is due to the
curvaton ``not knowing'' about the length of the inflationary period:
The curvaton field evolves and then starts to oscillate during inflation 
almost independently of the inflaton dynamics, especially for
small-field inflation.
Hence its energy density is almost uniquely determined by the e-folding
number, and as the curvaton dominates the universe it
dilutes away the consequences of the inflaton
fluctuations.

The field bounds (\ref{Lythphi}) and (\ref{Lythsigma}) now apply with 
$\kappa \simeq 3 \hat{r} / (4 + 3 \hat{r})$.
Since (\ref{Lythsigma}) no longer holds after the 
curvaton starts oscillating, one may think that the total field
excursion is reduced (in a similar fashion as in Footnote~[11]).
However an earlier onset of the oscillations
(i.e. larger~$\mathcal{N}_b$) quickly decreases~$\rho_\sigma$
and thus delays curvaton domination. 
One can check that $\mathcal{N}_b$ cannot be much bigger than $10$ for the
curvaton to dominate the universe before the Big Bang Nucleosynthesis,
hence the total $\sigma$-range is not greatly reduced. 

\begin{figure}[htbp]
 \includegraphics[width=.8\linewidth]{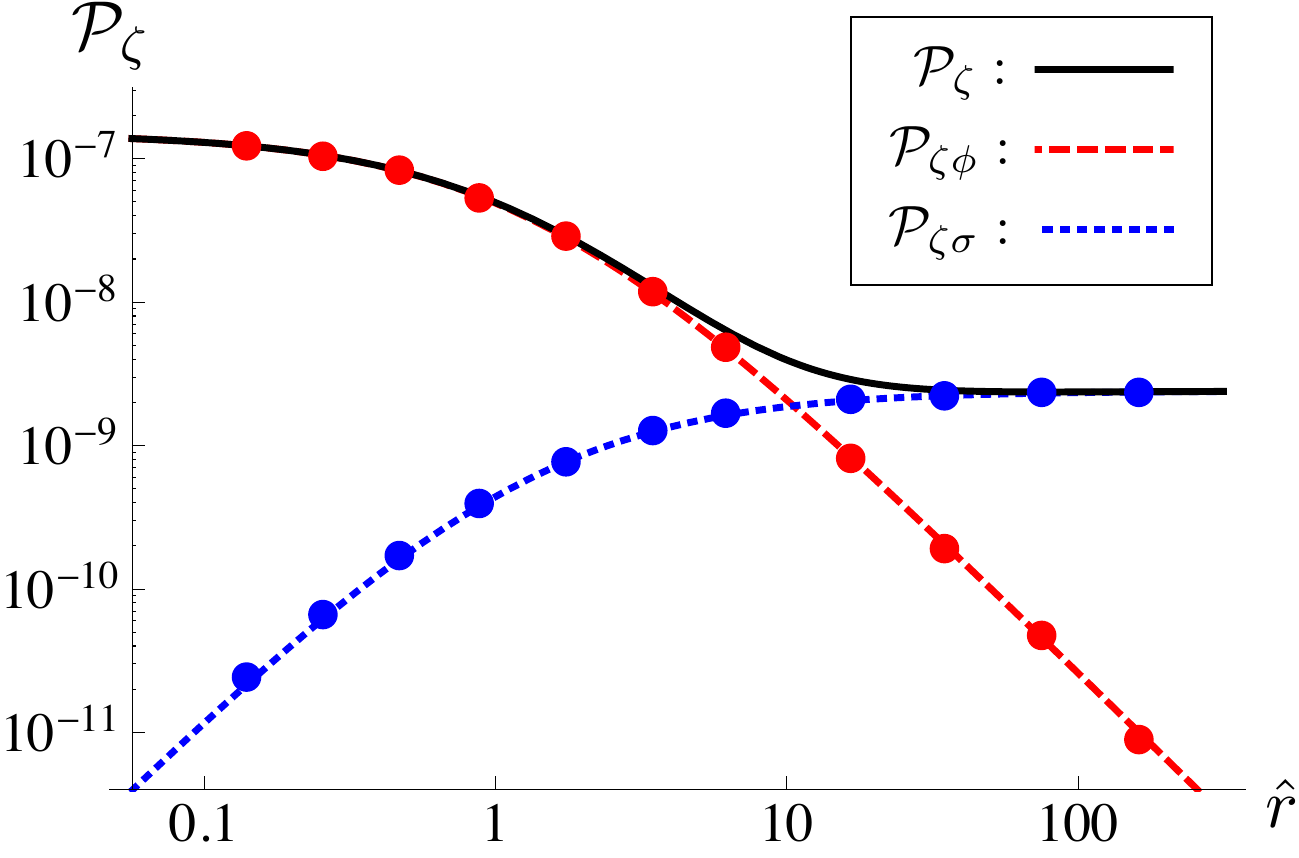}
 \caption{Amplitude of density perturbations as a function of the energy density
 ratio~$\hat{r}$. Inflaton-induced perturbations are suppressed at
 $\hat{r} \gg 1$.} 
 \label{fig:plot}
\end{figure}

We have carried out analytic and numerical calculations for 
a curvaton potential of the form
\begin{equation}
 V (\sigma) = \Lambda^4 \left(\frac{\sigma}{f}\right)^2 
 \left[ 1 + \left( \frac{\sigma}{f}\right)^2 \right]^{-1/2},
\label{linearpot}
\end{equation}
which is quadratic around the origin but approaches a linear potential for
$| \sigma | \gg f$. The curvaton initially slow-rolls along the linear
part, and then starts oscillating as it approaches $|\sigma| \sim f
$ \footnote{The initial oscillations along the linear part of the
potential may lead to formation of oscillons. 
However we do not expect them to modify the main results since they
behave as matter, and also because they should not affect
perturbations at the CMB scales which have exited the horizon
long before their formation.}.
The inflationary scale is fixed to
$H_{\mathrm{inf}} \approx 8.4 \times 10^{12}  \mathrm{GeV} $
and the curvaton potential tilt to
$(\Lambda^4 / f)^{1/3} \approx 1.8 \times 10^{14} \mathrm{GeV}$
so that when $\hat{r} \gg 1$, the curvaton-induced perturbations take the WMAP
value $\mathcal{P}_{\zeta \sigma} \approx 2.4 \times
10^{-9}$~\cite{Hinshaw:2012fq} and the tensor-to-scalar ratio is $r =
0.001$.
(The explicit values of $\Lambda$ and $f$ are irrelevant here, as long
as the effective mass $m_\sigma = \sqrt{2} 
\Lambda^2/f$ at the minimum is sufficiently larger than
$H_\mathrm{inf}$.)
We set the inflaton to possess a small-field type potential~$U(\phi)$ with an
almost constant tilt $\epsilon \equiv M_p^2 U'^2 / 2 U^2 \approx 10^{-6}$,
with field excursion $|\phi_* \! - \! \phi_{\mathrm{end}} |\approx 0.071 M_p$ 
to support about 50 e-folds between $t_*$ and $t_{\mathrm{end}}$. 
On the other hand, the initial curvaton field value is taken to be of
order the Planck scale $\sigma_* \sim 0.5 M_p $,
cf. (\ref{Lythsigma}). The resulting density
perturbations are plotted in Fig.~\ref{fig:plot} in terms of the
energy ratio~$\hat{r}$, which varies depending on the
decay rates $\Gamma_\phi$ and $\Gamma_\sigma$.
The analytic results from (\ref{Nphieo}) and (\ref{Nsigmaeo}) are shown
as lines (red dashed:  $\mathcal{P}_{\zeta \phi}$, blue dotted:
$\mathcal{P}_{\zeta \sigma}$, black solid: total~$\mathcal{P}_\zeta$),
while the dots represent numerical results 
(red:~$\mathcal{P}_{\zeta \phi}$, blue: $\mathcal{P}_{\zeta \sigma}$)
obtained by solving the fields' equations of motion and computing
$\delta \mathcal{N}$ by varying the initial field configuration.
The analytic and numerical results agree well, and 
one clearly sees that the inflaton-induced perturbations are suppressed
for $\hat{r} \gg 1$, allowing large tensor-to-scalar ratio with a small
inflaton field range. Note especially that $r$ is now much larger
than~$\epsilon$ due to the suppression of $\mathcal{P}_{\zeta \phi}$. 
Large energy ratio~$\hat{r}$ beyond the plotted range further suppresses
$\mathcal{P}_{\zeta \phi}$, until errors 
discussed below (\ref{Nsigmaeo}) become important. 
One can compute effects arising from a nonzero $\epsilon$ by solving the
field dynamics and check that $\partial \mathcal{N}/\partial \phi_*$
(22) further receives a contribution of  $\sim \epsilon \mathcal{N}_a \!
\times \! H_* / \dot{\phi}_*$ when $\hat{r} \gg 1$. 

\medskip
{\it Conclusions.}---
The density perturbations seeded by the inflaton can be
suppressed when having additional light fields.
This has important implications for inflationary cosmology, especially
when relating inflaton field excursions with the primordial gravitational waves.
Generalized forms of the Lyth bound (\ref{Lythphi}) and
(\ref{Lythsigma}) were also derived, showing that the field bound can be
evaded for the inflaton field but still holds for the total field space.
The results can further be extended to general multi-field inflation
models. 
We also note that the Lyth bound may be completely evaded when
having dynamical fields whose super-horizon fluctuations themselves are
suppressed, which may happen in some rapid-roll inflation models or
heavy oscillating curvaton scenarios. 

T.K. thanks Joel Meyers for helpful discussions. 
The work of T.T. is partially supported by the Grant-in-Aid for Scientific
research from the Ministry of Education, Science, Sports, and
Culture, Japan, No.~23740195.



\begin{thebibliography}{}

\bibitem{Lyth:1996im} 
  D.~H.~Lyth,
  Phys.\ Rev.\ Lett.\  {\bf 78}, 1861 (1997).

\bibitem{Baumann:2011ws} 
  D.~Baumann and D.~Green,
  JCAP {\bf 1205}, 017 (2012).

\bibitem{Linde:2005he} 
  A.~D.~Linde, V.~Mukhanov and M.~Sasaki,
  JCAP {\bf 0510}, 002 (2005).


\bibitem{Enqvist:2001zp}
  S.~Mollerach,
  Phys.\ Rev.\ D {\bf 42}, 313 (1990);
  A.~D.~Linde and V.~F.~Mukhanov,
  Phys.\ Rev.\ D {\bf 56}, 535 (1997);
K.~Enqvist and M.~S.~Sloth,
Nucl.\ Phys.\ B {\bf 626}, 395 (2002);
D.~H.~Lyth and D.~Wands,
Phys.\ Lett.\ B {\bf 524}, 5 (2002);
T.~Moroi and T.~Takahashi,
Phys.\ Lett.\ B {\bf 522}, 215 (2001)
[Erratum-ibid.\ B {\bf 539}, 303 (2002)].

\bibitem{Dvali:2003em}
  G.~Dvali, A.~Gruzinov and M.~Zaldarriaga,
  Phys.\ Rev.\  D {\bf 69}, 023505 (2004);
  L.~Kofman,
  arXiv:astro-ph/0303614.

\bibitem{Bernardeau:2002jf} 
  F.~Bernardeau and J.~-P.~Uzan,
  Phys.\ Rev.\ D {\bf 67}, 121301 (2003);
  D.~H.~Lyth,
  JCAP {\bf 0511}, 006 (2005).

\bibitem{BenDayan:2009kv} 
  I.~Ben-Dayan and R.~Brustein,
  JCAP {\bf 1009}, 007 (2010);
  Q.~Shafi and J.~R.~Wickman,
  Phys.\ Lett.\ B {\bf 696}, 438 (2011);
  S.~Hotchkiss, A.~Mazumdar and S.~Nadathur,
  JCAP {\bf 1202}, 008 (2012).


\bibitem{Kobayashi:toappear}
  A.~L.~Erickcek, N.~Kobayashi and T.~Kobayashi,
  in prep.

\bibitem{Kawasaki:2011pd} 
  M.~Kawasaki, T.~Kobayashi and F.~Takahashi,
  Phys.\ Rev.\ D {\bf 84}, 123506 (2011).
  
\bibitem{Hinshaw:2012fq} 
  G.~Hinshaw
  {\it et al.},
  arXiv:1212.5226 [astro-ph.CO].


\end{thebibliography}
\end{document}